# Zero-energy bound states in the high-temperature superconductors at the two-dimensional limit

Chaofei Liu[1†], Cheng Chen[1†], Xiaoqiang Liu[1], Ziqiao Wang[1], Yi Liu[1], Shusen Ye[1], Ziqiang Wang[2], Jiangping Hu[3,4,5,6] & Jian Wang[1,5,6,7]★

[1]International Center for Quantum Materials, School of Physics, Peking University, Beijing 100871, China
[2]Department of Physics, Boston College, Chestnut Hill, MA 02467, USA
[3]Beijing National Laboratory for Condensed Matter Physics & Institute of Physics, Chinese Academy of Sciences, Beijing 100190, China
[4]Kavli Institute of Theoretical Sciences, University of Chinese Academy of Sciences, Beijing 100190, China
[5]Collaborative Innovation Center of Quantum Matter, Beijing 100871, China
[6]CAS Center for Excellence in Topological Quantum Computation, University of Chinese Academy of Sciences, Beijing 100190, China
[7]Beijing Academy of Quantum Information Sciences, Beijing 100193, China

[†]These authors contributed equally to this work.
★E-mail: jianwangphysics@pku.edu.cn.

**Majorana zero modes (MZMs) that obey the non-Abelian statistics have been intensively investigated for potential applications in topological quantum computing. The prevailing signals in tunneling experiments "fingerprinting" the existence of MZMs are the zero-energy bound states (ZEBSs). However, nearly all of the previously reported ZEBSs showing signatures of the MZMs are observed in difficult-to-fabricate heterostructures at very low temperatures and additionally require applied magnetic field. Here, by using *in-situ* scanning tunneling spectroscopy, we detect the ZEBSs upon the interstitial Fe adatoms deposited on two different high-temperature superconducting one-unit-cell-thick iron chalcogenides on $SrTiO_3$(001). The spectroscopic results resemble the phenomenological characteristics of the MZMs inside the vortex cores of topological superconductors. Our experimental findings may extend the MZM explorations in connate topological superconductors towards an applicable temperature regime and down to the two-dimensional limit. While a concrete understanding of the observations is lacking, possible explanations involving novel 2D superconducting states with spin-orbit coupling, spontaneous nucleation of anomalous vortices at the magnetic sites, and noncoplanar magnetic ordering may further stimulate theoretical understandings of the scarcely captured ZEBSs in strongly correlated systems with multiband Cooper pairing.**

Quasiparticle excitations in superconductors crucially depend on impurity-scattering potentials, Cooper-pairing symmetry[1] and Andreev-reflection processes[2], etc. The zero-energy bound state (ZEBS) among them is particularly noteworthy, which can originate from the resonant tunneling involving topologically nontrivial Majorana zero mode (MZM)[3]. Obeying the non-Abelian braiding statistics, the MZM is a promising building block for the fault-tolerant topological quantum computation[4]. The theoretically predicted MZM platforms include Rashba 2D semiconductors[5], semiconducting nanowires[6], spin-textured Fe atomic chains[7] and topological-insulator ultrathin films[8] in proximity to Bardeen-Cooper-Schrieffer (BCS) superconductors. The ZEBSs proposed as the signatures of MZMs in these systems have been experimentally detected[9-12]. However, the heterostructure-fabricating difficulty, the low experimental temperature (primarily ≤ 1.4 K) and the extra requirement of an applied magnetic field for those designs make the future applications of MZMs in quantum-functionality electronics highly challenging.

The iron-based high-temperature superconductors provide alternative directions for pursuing the MZMs, where the topologically nontrivial phases have been theoretically predicted to universally exist[13-18]. In recent experiments, the simultaneously discovered Dirac-type spin-helical surface state and *s*-wave superconducting (SC) gap in the iron-chalcogenide material, $FeTe_{0.55}Se_{0.45}$, highlight the existence of connate topological superconductivity at high temperatures[19]. According to the Fu-Kane model, the proximitized interface between topological surface state (TSS) and *s*-wave pairing potential resembles a spinless $p_x+ip_y$ superconductor, which can host the MZMs in SC-order-parameter defects[8]. Consistently, the ZEBSs bearing the characteristics of MZMs have been observed at the interstitial iron impurities (IFIs) and a fraction of magnetic vortices of bulk superconducting Fe(Te,Se) single crystals[20,21]. Although bulk Fe(Te,Se) is a nominally perceived high-temperature superconductor, its SC transition temperature ($T_c$) is limited below 15 K[20]. The relatively low $T_c$ of Fe(Te,Se), together with the difficult-to-control character of magnetic-field–induced vortices therein, poses barriers to technically realizing and freely manipulating the MZMs. It is accordingly desirable to investigate the Majorana physics both in higher-$T_c$ superconductors and under more-feasibly-manipulable physical settings.

The one-unit-cell (1-UC) FeSe[22] and Fe(Te,Se)[23] on SrTiO$_3$(001), showing an enhanced high $T_c$ typically above 60 K[23-25] and predicted to be topologically nontrivial[16,18], exemplifies potential platforms to address those issues. Furthermore, although the magnetically ordered Yu-Shiba-Rusinov (YSR) (magnetic) chains placed on BCS superconductors have been systematically investigated in connection to MZMs[7], the individual-YSR-atom–based MZM configurations are scarcely proposed. Here, by *in-situ* scanning tunneling spectroscopy (STS) (4.2 K unless specified; see Methods), we report the experimental discovery of the MZM-like ZEBSs induced by interstitial Fe adatoms on high-quality 1-UC FeSe and FeTe$_{0.5}$Se$_{0.5}$ on SrTiO$_3$(001) in the absence of external magnetic field. The spectroscopic results establish the generic existence of the ZEBSs in iron-chalcogenide high-temperature superconductors at the two-dimensional (2D) limit.

In the Brillouin zone (BZ) of 1-UC FeSe/SrTiO$_3$(001) (Fig. 1A), the Fermi surface consists of only electron pockets at the M points (Fig. 1, B and C)[25]. By using the molecular-beam-epitaxy (MBE) technique, our 1-UC FeSe films were well prepared with atomically flat surfaces at both mesoscopic and microscopic scales (Fig. 1, D and E). As in previous reports[22,25], the tunneling spectrum (d$I$/d$V$ vs. $V$) measured on 1-UC FeSe surface is fully gapped and typifies the multiband superconductivity as suggested by the high-resolution photoemission spectroscopy (Fig. 1F)[26]. Along the trajectory in a single domain, the spatially resolved spectra repeatedly reveal the multiband pairing state with SC gaps near 9.5 and 16.0 meV, respectively (Fig. 1G). Therefore, a pristine high-temperature SC substrate [1-UC FeSe/SrTiO$_3$(001)] for the adsorbate deposition is well established.

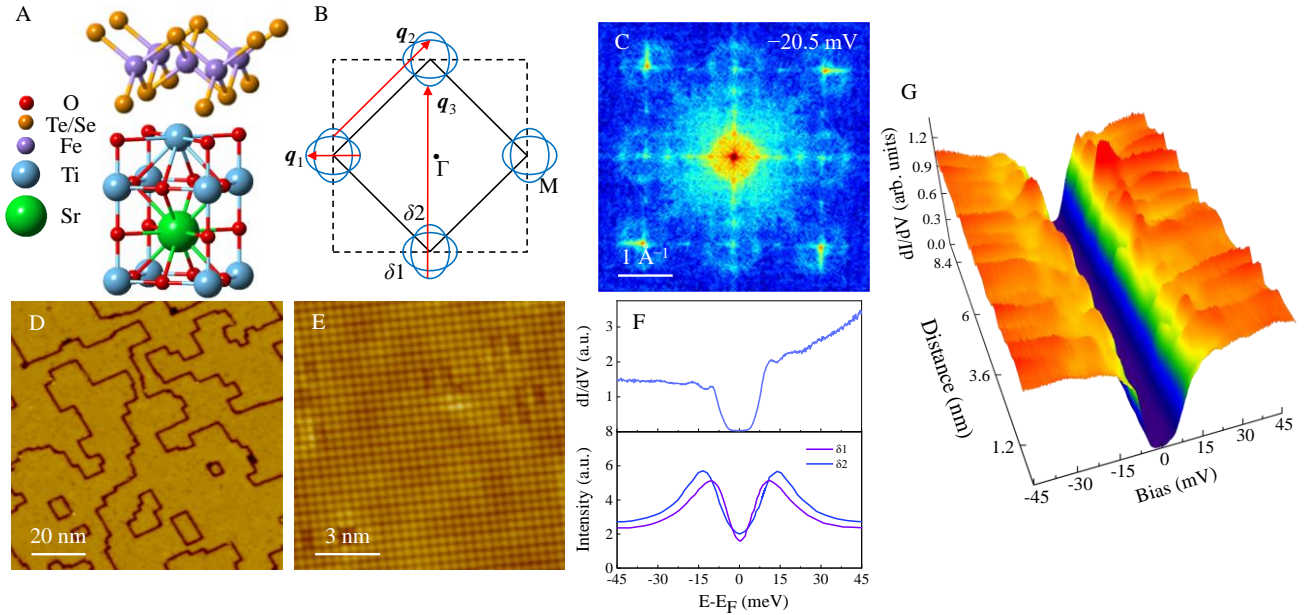

**Fig. 1. Crystal structure, topographies and multiband superconductivity of 1-UC FeSe/SrTiO$_3$(001).** (A) Crystal structure and (B) Fermi-surface topology in the folded BZ (solid black square) of 1-UC FeSe/SrTiO$_3$(001). $q_1$, $q_2$ and $q_3$: possible scattering vectors. (C) Fast Fourier transformation of the d$I$/d$V$ mapping at −20.5 mV measured on pristine 1-UC FeSe surface (28×28 nm$^2$; set point: $V$ = 0.06 V, $I$ = 500 pA; modulation: $V_{mod}$ = 1 mV). Pocket-like scattering vectors can be seen and coincide with the hole-pocket–absent Fermi-surface topology in (B). (D,E) Topographic images of 1-UC FeSe

film at different scales [D: 100×100 nm$^2$, E: 12.5×12.5 nm$^2$; set point: (D) $V$ = 0.6 V, $I$ = 500 pA, (E) $V$ = 0.2 V, $I$ = 500 pA]. (F) Comparison between tunneling spectrum (upper panel) and angle-resolved photoemission spectra (lower panel)[26] of 1-UC FeSe. The positions of $\delta$1 and $\delta$2 are marked in (B). (G) Spatially resolved tunneling spectra along a 9.4-nm trajectory. (F,G) Set point: $V$ = 0.04 V, $I$ = 2500 pA; modulation: $V_{mod}$ = 1 mV (by default unless specified).

The Fe atoms were deposited on the 1-UC FeSe surface at ~143–155 K at an ultralow coverage (Supplementary Part I) for the formation of individual adatoms (Fig. 2A). In space, the Fe adatom locates above the energetically favorable, highly symmetric interstitial hollow site of four adjacent Se atoms (see also Fig. S1A). Spectroscopically, the ZEBS modulated by adsorbate-substrate interaction (Supplementary Part I) is induced by the interstitial Fe adatom, which appears as a zero-bias conductance peak (ZBCP) in the tunneling spectrum and is exceptionally sharp with a peak-to-dip d$I$/d$V$ ratio of ~3 (Fig. 2B). Both the ultrahigh crystalline quality of 1-UC FeSe (Fig. 1, D and E) and the well-defined isolation of the adatom (Fig. 2A) exclude the extrinsic disorder effect on the detected ZBCP signal. Even sharper ZBCP lineshape is expected when cooling down the experimental temperature (4.2 K) far below 1 K by a $^3$He or a dilution refrigerator, which is beyond our current equipment capability and would stimulate further investigations in this direction. The magnetic-impurity–induced ZEBS has been rarely reported in fully gapped superconductors[20] and constitutes the major finding in current study. Also note that, the Fe adatom shows no sign of the ZEBS on ~30-UC FeSe/SiC(0001)[27]. The contrast of the ZEBS that is present in 1-UC FeSe/SrTiO$_3$ and absent in ~30-UC FeSe/SiC, respectively, signifies the necessity of topological nontrivial phases predicted only in the former[18] for the emergence of ZEBS.

To deeply reveal the ZEBS properties, the isolated interstitial Fe adatom was investigated in detail (Fig. 2, C–E). The spatial evolution of the tunneling spectra along a linecut departing from the adatom is presented in Fig. 2C. As moving away from the adatom center, the zero-bias signal drops abruptly but remains a single peak before becoming unidentifiable (see also Fig. S2, A and B). The unsplitting behavior of the ZEBS here is noteworthy and reminiscent of the unsplit MZM-like ZEBS off magnetic-vortex center in SC TSS[21]. Moreover, the integrated low-bias density of states keeps roughly constant for the line spectra in Fig. 2C (Fig. S2C), implying a spectral-weight transfer from the coherence peaks to the ZBCP. To directly visualize the ZEBS distribution in space, a d$I$/d$V$ mapping for the Fe-adatom topography in Fig. 2A was measured at 0 mV (Fig. 2D). Enhanced feature intimately bounded to the adatom edge was found in the ZEBS pattern, which correlates with the phase decoherence by the Fe adatom. For a more quantitative analysis, the linecut profiles starting from the adatom center were extracted from Fig. 2D and one of them, L, is exemplified in Fig. 2E. The exponential fitting of L yields a decay length, $\xi$, of 3.4 Å, which is nearly one order of magnitude smaller than the SC coherence length (2.45 nm)[28].

Kondo resonance[29] and impurity-scattering state[1] are available as the possible physical origins of the detected ZEBS, both of which can appear as single peaks around zero bias in the tunneling spectra. The Kondo effect originates from the exchange interaction between localized spins of magnetic impurities and conducting electrons in simple metals (Ag, Au, Cu, Pb etc)[30]. In experiments, the Kondo resonance is well described by a Fano lineshape[29] and should deviate from zero energy due to finite potential scatterings in the SC state[1]. Thus, the non-Fano–like (strikingly sharp) and precisely 0-mV conductance peak here is at odds with the Kondo scenario. To put further constrains to the interpretations for the detected ZEBS, temperature-dependent experiment for the adatom-induced ZBCP was performed for 1-UC FeSe (Fig. 3A). As the temperature increases, the peak intensity in the experimental spectrum decreases and completely merges into the background at 13 K well below $T_c$. Hence, the ZEBS is an emergent "sub-SC" phenomenon, where the nodeless SC gap away from the adatoms can be fully reserved (Fig. 2, B and C). If the ZEBS feature is dominated by the Kondo resonance, the ZBCP shall be insensitive to superconductivity and persist far above $T_c$[31], which is evidently inconsistent with the observation here (Fig. 3A). On the other hand, being mutually correlated, the impurity state and the superconductivity should exhibit synchronized temperature dependence by the thermal-broadening effect[32]. The temperature evolution of the ZBCP with assumed impurity-state origin is obtained by convoluting the 4.2-K spectrum taken upon the Fe adatom by higher-temperature Fermi-Dirac distribution function (Fig. 3A). Apparently, the zero-bias conductances (ZBCs) ($G_p$) as a function of temperature for the experimental and convoluted ZEBS spectra substantially deviate from each other (Fig. 3B), quantitatively indicating the conventional impurity-state explanation for the ZEBS is incompatible. Especially, in previous experimental studies, the near-zero-energy impurity states were mainly observed in $d$-wave superconductors at the strong-scattering (unitary) limit[1]. Besides the non–$d$-wave [e.g., extended $s_\pm$-wave[33]] pairing possibility of the 1-UC FeSe/SrTiO$_3$(001), the scattering from the Fe adatom inducing the ZEBS here is not at the unitary limit in statistics (Supplementary Part I).

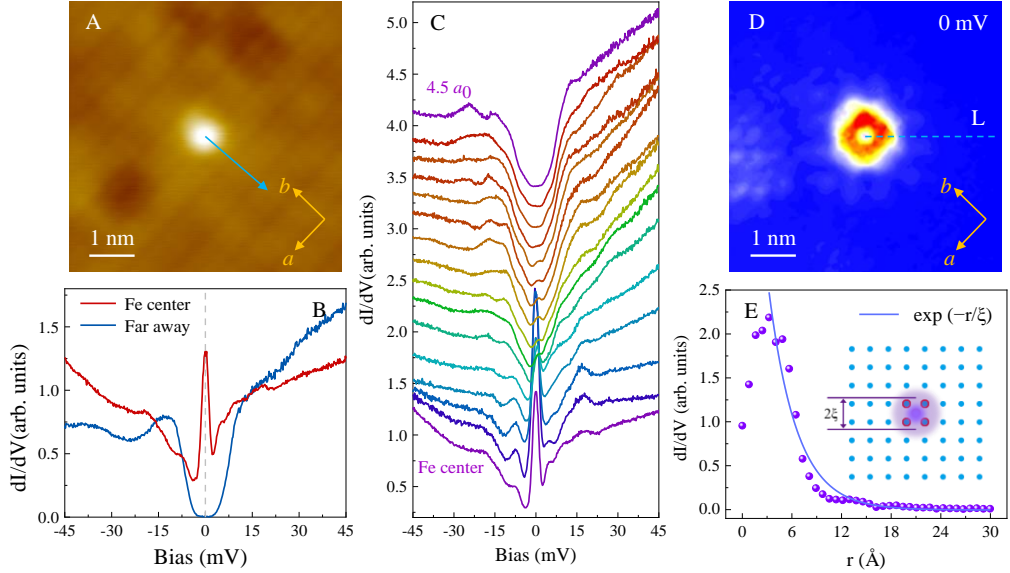

**Fig. 2. Spatial evolution of the ZEBS in 1-UC FeSe.** (A) Topographic image of an isolated Fe adatom (6×6 nm$^2$; set point: $V$ = 0.2 V, $I$ = 500 pA). (B) Tunneling spectra taken upon the Fe-adatom center and far away. (C) Spatially resolved tunneling spectra (vertically offset for clarity) along the arrow in (A). (D) d$I$/d$V$ mapping at 0 mV for the Fe adatom in (A). (E) Linecut (solid symbols) along the dashed line, L, in (D) and corresponding exponential fitting (solid curve), showing ZBC as a function of the distance, $r$, relative to the Fe-adatom center. Fitting formula: d$I$/d$V$(r, 0 mV) ∝ exp(−$r/\xi$); $\xi$: decay length. Inset: schematic of the spatial distribution of Fe-adatom scattering.

Recent theoretical advances have extended the magnetic-impurity states to iron-based superconductors treated as spin-singlet pairing with both sign-preserving ($s_{++}$-wave) and sign-reversing ($s_\pm$-wave; nodal and nodeless $d$-wave) SC-gap functions over the Fermi surfaces[34-37]. In these models, the impurity-state energy can be tuned to zero for a critical strength of the scattering potential[35,37], which appears in agreement with the detected ZEBS and demands more experiments to scrutinize the ZEBS identity. Physically, the impurity state is doubly degenerate at zero energy with both spin-up and spin-down components[31], which will be split by an applied magnetic field. This is analogous to the theoretical prediction that the quantum-mechanically coupled magnetic-adatom dimer will hybridize and split the degenerate impurity state into bonding and antibonding counterparts[38], where one adatom is effectively situated in the magnetic-exchange field of the other. The unavailability of external magnetic field in our apparatus motivates us to alternatively study the ZEBS response in the adatom dimer. Surprisingly, the ZEBS remains a robust single peak at zero energy for two closely located interstitial Fe adatoms on 1-UC FeSe (Fig. 3, D and E). The presence of spin-orbit coupling (SOC) has been shown to suppress the Zeeman-field–induced splitting of the ZEBSs localized at magnetic impurities for bulk Fe(Te,Se) with electron and hole Fermi-surface nesting and $s_\pm$-type pairing[39]. While whether this theoretical conclusion is valid remains to be examined for the Fe-adatom dimer in 1-UC FeSe, where the hole Fermi pocket is absent in contrast, all these aforementioned results for the detected ZEBS in 1-UC FeSe are hardly reconciled within the conventional impurity-state framework.

The inconsistency between ZBCP and nontopological interpretations raises the concern that whether a topologically nontrivial picture is responsible for the ZEBS emergence. The 2D superconductors with Rashba SOC can turn into the topological superconducting (TSC) states via sufficiently strong time–reversal-symmetry–breaking Zeeman field, where the MZMs emerge as the ZEBSs in magnetic-vortex cores. Equivalently, the exchange field from generic magnetic impurities can serve in place of the Zeeman coupling to break the time-reversal symmetry. The physical properties of the magnetic-impurity–induced YSR states in this type of topological superconductors and superfluids, including bound-state spectra, band structures, wavefunctions and spatial profiles of SC order parameters, have been extensively investigated and found to be intriguingly different from those in nontopological cases[40-43]. Recently, the quantum anomalous vortices (QAVs) were theoretically proposed to be generated by interstitial magnetic transition-metal ions in a strongly spin–orbit-coupled *s*-wave superconductor in the absence of external magnetic field[44]. When proximity-coupled to the spin-helical Dirac-fermion TSS, the magnetic-ion–trapped QAV supports the MZM in the core, which follows the same spirit of the Fu-Kane model[8] but without requiring the external field. There is surely no TSS available in connate 1-UC FeSe. However, if the 2D SC state contains the necessary ingredients for being topologically nontrivial, the nucleation of the QAV at the interstitial Fe-adatom site is likely to induce a ZEBS-typified MZM. Indeed, mainly based on the observation of Fermi-level–crossed linearly dispersive bands, the undoped parent phase of 1-UC FeSe/SrTiO$_3$ was demonstrated to be Dirac-semimetallic in nature by photoemission spectroscopy[45], which can be induced by the substrate tensile-strain field and SOC[17]. When electron doping is introduced, an odd-parity TSC state was theoretically conjectured to emerge in the nodeless *s*-wave pairing channel[18]. While much work is needed to explore the topological properties of the SC state in 1-UC FeSe, these recent developments, together with the detected ZEBS and its incompatibility especially with the conventional impurity-scattering state, suggest the QAV to be a possible candidate for the ZEBS physics. We are thus motivated to further investigate the characteristics of the ZBCP in 1-UC FeSe, and compare them with those of the MZM that behaves as the ZEBS within the topologically nontrivial scenario.

Phenomenologically, the ZBCP-"melting" temperature ($T_c^{ZBCP}$) in nearly all literatures reporting the MZM-like signatures[9,10,21,46] is far below the $T_c$ of the pristine SC components in corresponding MZM configurations. This empirical relation, $T_c^{ZBCP} \ll T_c$, is well satisfied in the present 1-UC FeSe situation (i.e., ~13 K ≪ ~60 K), implying the likelihood of Majorana signal involved in the Fe-adatom–induced ZEBS. By comparing with the impurity-state–like convoluted spectra in Fig. 3A, where the ZBCP persists up to ≥35 K, we conclude that the disappearance of the experimental ZBCP above 13 K is not a simple consequence of thermal broadening. Thus, additional $T_c^{ZBCP}$-suppression mechanism must exist, which can be described by the Majorana "poisoning" through thermally excited quasiparticles[47].

The full width at half maximum (FWHM) of the MZM peak in the tunneling spectrum is determined by instrumental, thermal and tunneling broadening[48]. To quantitatively test the MZM explanation for the ZEBS, the FWHM of the ZBCP in the experimental and convoluted spectra at various temperatures (Fig. 3A) was extracted and shown in Fig. 3C. The spectrum broadening, $\Delta E$, combining both instrumental and thermal effects, is also plotted for comparison. It is seen that the FWHM of the ZBCP is limited by $\Delta E$ for the experimental ZEBS spectra, falling within the MZM picture, but exceeds $\Delta E$ for the convoluted ZEBS spectra for temperature ≥ 13 K. Moreover, as shown in Fig. 3F, the experimental ZEBS at 4.2 K exists robustly and the ZBCP FWHM remains approximately $\Delta E$-limited across nearly three orders of magnitude in the tunneling-barrier conductance $G_N$. The fact that the experimental ZEBS is mainly spectrum-broadened by $\Delta E$ indicates a negligible tunneling-coupling strength ($\Gamma$) and especially coincides with the spectral expectation for the Majorana ZBCP of an intrinsically single-peak nature. The latter excludes the possibilities of both impurity-state–related multifold degeneracy at zero energy and densely packed low-energy quasiparticle excitations around Fermi level. That only the ZEBS exists within the SC gap is consistent with the expulsion of nonzero-energy Caroli-de Gennes-Martricon (CdGM) vortex-core states beyond the SC gap in the QAV theory, protecting the MZM by preventing the mixing with topologically trivial higher-level CdGMs[44].

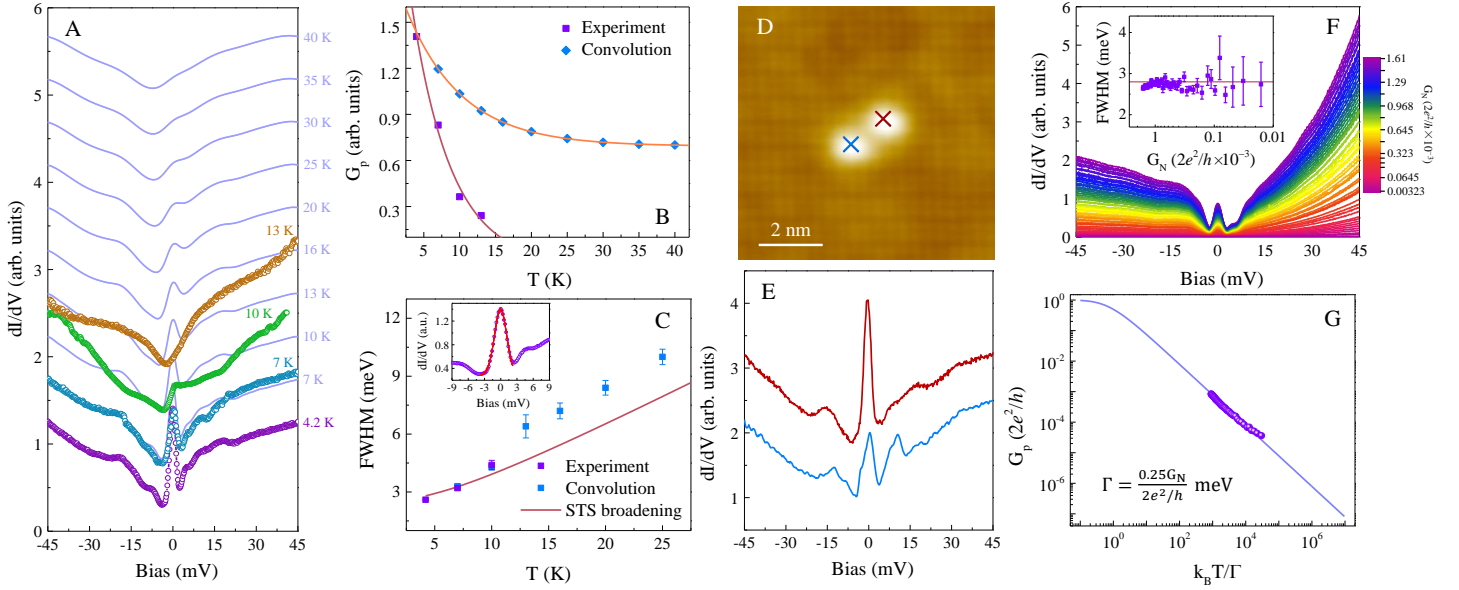

**Fig. 3. Perturbation of the ZEBS in 1-UC FeSe by temperature, a neighboring Fe adatom and different tunneling barriers.** (A) Temperature dependence of the experimental tunneling spectra (open symbols) measured upon the Fe-adatom center in Fig. 2A and the convoluted 4.2-K spectra (solid curves) by Fermi-Dirac distribution function at higher temperatures (both vertically offset for clarity). (B) ZBC, $G_p$, plotted as a function of temperature (solid symbols) extracted from (A). (C) FWHM of the ZBCP (solid symbols) in the experimental and convoluted spectra of (A). The solid curve is the spectral energy resolution combining both instrumental and thermal broadening, $\Delta E = \sqrt{(2.5eV_{\mathrm{mod}})^2 + (3.5k_\mathrm{B}T)^2}$. Inset: exemplifying the Gaussian fitting to the ZBCP, which defines the FWHM and error bars (standard deviations) adopted throughout. (D) Topographic image of an Fe-adatom dimer (8×8 nm²; set point: $V$ = 0.1 V, $I$ = 500 pA). (E) Tunneling spectra (vertically offset for clarity) taken upon the Fe adatoms in (D). (F) Tunneling-barrier-conductance $G_N$ dependence of the ZEBS spectra (4.2 K), where $G_N$ is parameterized as the tunneling conductance ($G_N = I/V$). Inset: FWHM of the ZBCP in the main-panel spectra under different $G_N$. The solid curve is $\Delta E$ at 4.2 K. (G) Scaling analysis of $G_p$ of the tunneling spectra of (F). Solid curve: calculated $G_p$ for the MZM at different $k_\mathrm{B}T/\Gamma$ according to Eq. 1; solid symbols: experimental $G_p$ plotted as a function of $G_N$ rescaled as $k_\mathrm{B}T/\Gamma$ to match the calculated curve, yielding $\Gamma = \frac{0.25 G_N}{2e^2/h}$ meV.

Dictated by particle-hole symmetry, the incoming electron and reflected hole in the tunneling process into a MZM via resonant Andreev reflection equally contribute to the zero-energy tunneling amplitude, resulting in the quantized ZBC of $2e^2/h$ at zero temperature[3]. Because of the small tunneling-coupling strength Γ and the Majorana poisoning by nontopological quasiparticles, the ZBC quantization is extremely challenging to achieve in STS experiments. Nevertheless, due to the universal scaling function obeyed by Majorana ZBCP[49], the ZBCs below conductance-quantization region can be utilized to trace down to the zero-temperature limit as a consistency check of the ZEBS nature. Specifically, at weak-tunneling ($\Gamma \ll \Delta_{1,2}$) and low-temperature ($k_\mathrm{B}T \ll \Delta_{1,2}$; $k_\mathrm{B}$: Boltzmann constant) conditions, the ZBC of tunneling-broadened finite-temperature MZM spectrum in subgap regime, $G_p$, is

given by

$$G_p = \frac{2e^2}{h} \int_{-\infty}^{\infty} \frac{\Gamma^2}{E^2+\Gamma^2} \frac{1}{4k_BT\cosh^2(E/2k_BT)} dE, \quad (1)$$

which only depends on the dimensionless ratio, $k_BT/\Gamma$[50,51]. In accordance with the spinless model for MZM tunneling, $\Gamma = \eta \frac{G_N}{2e^2/h}$ for sufficiently small $\frac{G_N}{2e^2/h}$[48], corresponding to the situation in STS configurations. The $G_N$-dependent $G_p$ extracted from Fig. 3F for 1-UC FeSe can be rescaled as a function of $k_BT/\Gamma$ at different values of $\eta$, which is found to collapse onto the scaling curve (Eq. 1) for $\eta = 0.25$ (Fig. 3G). The yielded $\Gamma = \frac{0.25 G_N}{2e^2/h}$ meV for our experimental $G_N$ range is ~4–6 orders of magnitude smaller than the instrumental and thermal spectrum broadening, $\Delta E$. The vanishingly small $\Gamma$ is reasonably consistent with the weak-tunneling condition as concluded from the ZEBS-spectrum-linewidth analysis and naturally explains the weak $G_N$ dependence of ZBCP FWHM (Fig. 3F). All above experimental observations resemble phenomenologically the spectroscopic signatures of the MZM. Pushing the experimental data [$G_p(G_N)$] towards the asymptotic small-$k_BT/\Gamma$ limit at ultralow temperatures and ultrahigh tunneling conductance will unveil more solid evidence for the MZM in future investigations.

The sharply defined ZBCP induced by the interstitial Fe adatom was reproduced in a different 2D iron-chalcogenide high-temperature superconductor, 1-UC FeTe$_{1-x}$Se$_x$ ($x \approx 0.5$) (Fig. 4, A and B). The preparation of high-quality 1-UC Fe(Te,Se) film showing high $T_c$ is exceedingly challenging experimentally and the *in-situ* STS studies have been rarely reported. By utilizing the MBE technique (Methods), we successfully grew the highly disorder-free 1-UC FeTe$_{0.5}$Se$_{0.5}$ on SrTiO$_3$(001) with both mesoscopic and microscopic atomic flatness. The nominal composition of 1-UC FeTe$_{0.5}$Se$_{0.5}$ was calculated by measuring the thickness of the 2$^{nd}$-UC Fe(Te,Se) film and independently counting the ratio of Se/Te. Both methods yield consistent stoichiometry of 1-UC Fe(Te,Se). As observed in 1-UC FeSe, the ZEBS in 1-UC FeTe$_{0.5}$Se$_{0.5}$ induced by the interstitial Fe adatom maintains an unsplit ZBCP spectrum that fades out spatially away from the adatom (Fig. 4C) and is intimately localized by the adatom (Fig. 4D) with a decay length of roughly 6.4 Å (Fig. 4E).

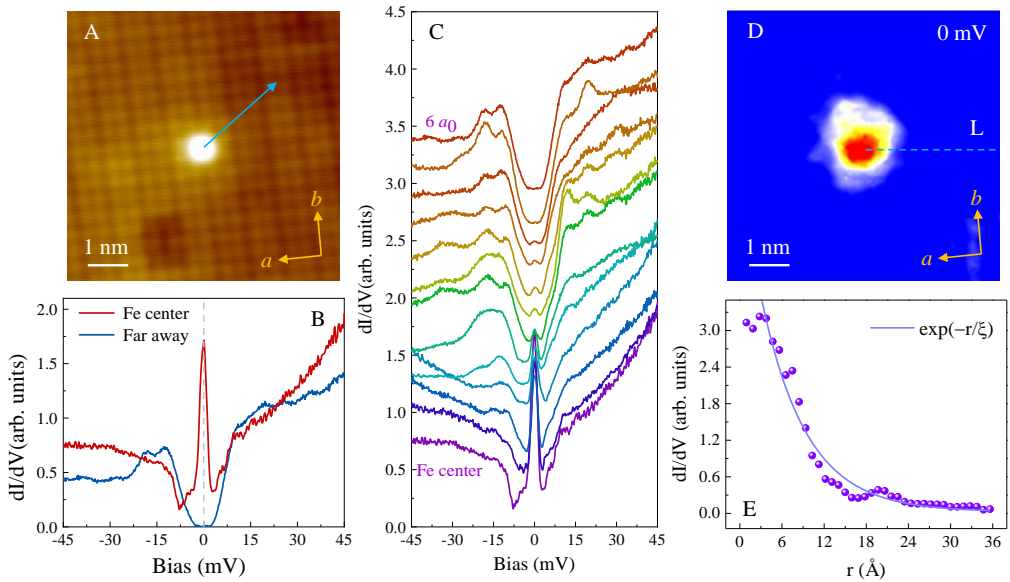

**Fig. 4. Spatial evolution of the ZEBS in high-quality 1-UC FeTe$_{0.5}$Se$_{0.5}$/SrTiO$_3$(001).** (A) Topographic image of an isolated Fe adatom (6.75×6.75 nm$^2$; set point: $V$ = 0.1 V, $I$ = 500 pA). (B) Tunneling spectra taken upon the Fe-adatom center and far away. (C) Spatially resolved tunneling spectra (vertically offset for clarity) along the arrow in (A). (D) d$I$/d$V$ mapping at 0 mV for the Fe adatom in (A). (E) Linecut (solid symbols) along the dashed line, L, in (D) and corresponding

exponential fitting (solid curve) by $dI/dV(r, 0\,mV) \propto \exp(-r/\xi)$, showing ZBC as a function of the distance, $r$, relative to the Fe-adatom center.

The spectroscopic characteristics of the ZEBS in 1-UC FeSe are all highly reproducible in 1-UC FeTe$_{0.5}$Se$_{0.5}$ (Fig. 5). Compared to 1-UC FeSe, the ZBCP in the experimental spectrum of 1-UC FeTe$_{0.5}$Se$_{0.5}$ disappears at a higher temperature of 20 K which is still well below $T_c$, in contrast to the thermally convoluted ZEBS spectrum with assumed impurity-state origin (Fig. 5, A and B). For an Fe-adatom dimer, the ZEBS spectrum remains singly peaked (Fig. 5, D and E). Both the premature thermal melting and the spectral unsplitting against local magnetic-exchange field for the detected ZEBS in 1-UC FeTe$_{0.5}$Se$_{0.5}$ reemphasize the difficulty in describing the ZEBS in terms of the conventional impurity-scattering state. Furthermore, the FWHM result is consistent with the intrinsically single-peak nature of the detected ZBCP (Fig. 5C) as expected of the MZM. Most intriguingly, the experimentally detected ZEBS (4.2 K) exhibits rather robust existence for a wide range of the tunneling-barrier conductance $G_N$ over several orders of magnitude (Fig. 5F). By detailed scaling analysis based on Eq. 1, the $G_N$-dependent $G_p$ extracted from the experimental ZEBS spectra in Fig. 5F is quantitatively described by the universal scaling of Majorana ZBCP (Fig. 5G), yielding $\eta = 0.35$ comparable with that (0.25) obtained in 1-UC FeSe.

By first-principles calculations, FeTe$_{1-x}$Se$_x$ ($x = 0.5$) has been predicted to show the nontrivial topology characterized by an odd $Z_2$ invariant[15], which can survive down to the 1-UC limit for $x < 0.7$[16]. Therefore, the epitaxially prepared 1-UC FeTe$_{0.5}$Se$_{0.5}$ is situated in the predicted topological regime of both bulk and ultrathin FeTe$_{1-x}$Se$_x$. It is truly remarkable that the robust MZM-like ZEBSs persist at the interstitial Fe-adatom sites in the 1-UC FeTe$_{0.5}$Se$_{0.5}$ as shown by our experiments. The reproducibility of the phenomenological MZM features in different 1-UC iron chalcogenides possibly suggests a common topologically nontrivial origin of the detected ZEBSs.

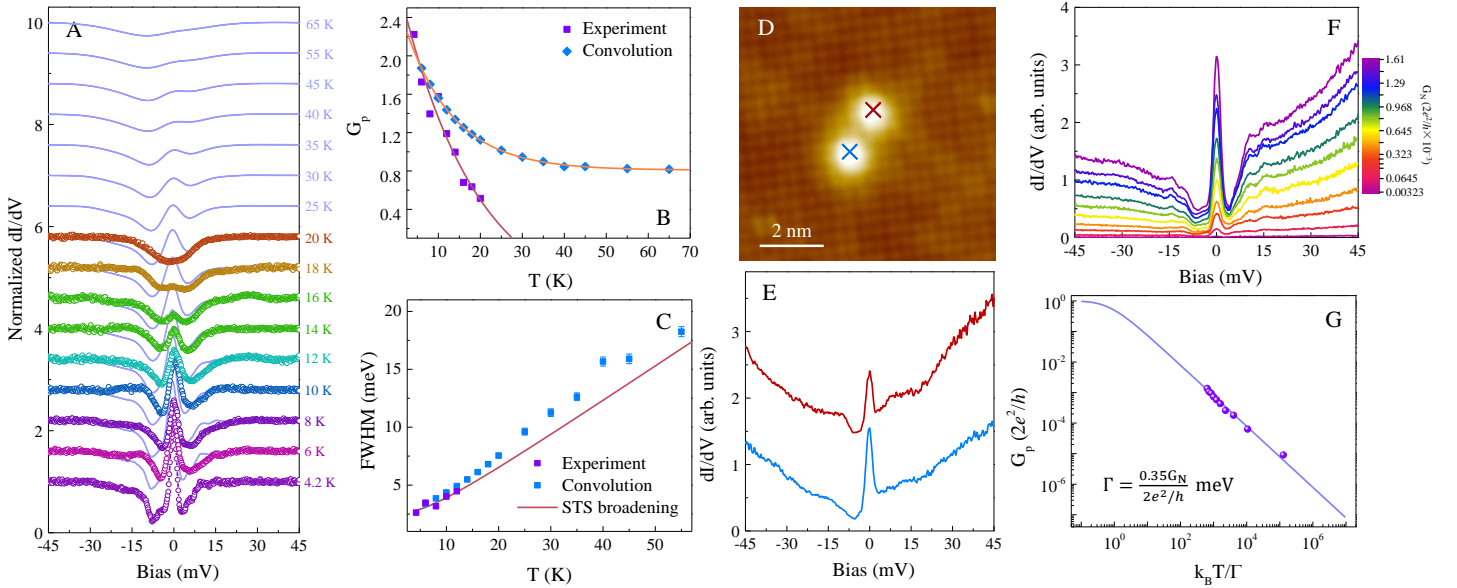

**Fig. 5. Perturbation of the ZEBS in 1-UC FeTe$_{0.5}$Se$_{0.5}$ by temperature, a neighboring Fe adatom and different tunneling barriers.** (A) Temperature dependence of the experimental tunneling spectra (open symbols) measured upon the Fe-adatom center in Fig. 4A and the convoluted 4.2-K spectra (solid curves) by Fermi-Dirac distribution function at higher temperatures (both vertically offset for clarity). The experimental spectra have been normalized by their respective cubic-polynomial backgrounds for clarity, which are obtained from fitting to the spectra for bias $|V| \geq 30$ mV. (B) $G_p$ plotted as a function of temperature (solid symbols) extracted from (A). (C) FWHM of the ZBCP (solid symbols) in the experimental and convoluted spectra of (A). The solid curve is the spectral energy resolution combining both instrumental and thermal broadening ΔE plotted vs. temperature. (D) Topographic image of an Fe-adatom dimer (8×8 nm$^2$; set point: $V$ = 0.1 V, $I$ = 500 pA). (E) Tunneling spectra (vertically offset for clarity) taken upon the Fe adatoms in (D). (F) $G_N$ dependence of the ZEBS spectra (4.2 K). (G) Scaling analysis of $G_p$ of the tunneling spectra of (F). Solid curve: calculated $G_p$ for the MZM at different $k_BT/\Gamma$ according to Eq. 1; solid symbols: experimental $G_p$ plotted as a function of $G_N$ rescaled as $k_BT/\Gamma$ to match the calculated curve, yielding $\Gamma = \frac{0.35 G_N}{2e^2/h}$ meV.

In summary, we have reported the observation of the ZEBSs manifested as the ZBCPs in STS upon the interstitial Fe adatoms on two different 2D high-temperature superconductors, 1-UC FeSe and FeTe$_{0.5}$Se$_{0.5}$. All the presented data are highly similar for these two compounds, both demonstrating the inadequacy of the ZEBS interpretation based on Kondo resonance and conventional impurity state. More detailed experiments of the ZEBSs show overall consistency with the MZM phenomenology, including their temperature and tunnel-barrier-conductance dependences. The analyses of the ZBCP FWHM and the tunneling coupling suggest the spectroscopic absence of other adatom-induced bound states, possibly expelled outside the SC gap due to the exchange field of the Fe adatoms. This situation is analogous to the theoretical proposal of the QAVs nucleated at the magnetic ions in superconductors with strong SOC, where the exchange field leads to the expulsion of the CdGM vortex-core states beyond the SC-gap energy[44]. In light of the possible topological phases[16-18,45] and SOC-induced spin-triplet pairing components in 1-UC FeSe and Fe(Te,Se), the hidden mechanism responsible for the observed ZEBSs may be the QAVs nucleated at the magnetic Fe adatoms in 2D topological superconductors. Different from the Fu-Kane model[8], within the QAV scenario for the MZMs in the ultra-2D 1-UC FeSe and FeTe$_{0.5}$Se$_{0.5}$ in the absence of external magnetic field, the magnetic-flux lines produced by and connecting the quantum anomalous vortex-antivortex pairs are expected to form closed loops around the 1-UC SC layer[44].

Other possibilities for the topological origin of the ZEBS include the existence of helical topological mirror superconductivity[41], odd-frequency pairing state[52], $d+id$-wave superconductivity[53], and the potentially modulated magnetic skyrmion[54] preluded by Fe-adatom–reconstructed electronic structures (Supplementary Part III). We emphasize that the observed ZEBSs in truly high-temperature superconductors at the 2D limit are completely new and existing theories cannot provide a precise and direct explanation at the present time. Here, we have discussed just a few possibilities and hope our findings will stimulate further theoretical and experimental investigations. A thorough understanding of the observed ZEBSs would require detailed knowledge of the interplay among adsorbate-substrate interaction, local magnetic structure, SOC and superconductivity.

Compared to bulk Fe(Te,Se) studied in refs 20,21, the high $T_c$ of 1-UC FeSe and FeTe$_{0.5}$Se$_{0.5}$ significantly increases $T_c^{ZBCP}$ (Supplementary Part IV), which is an essential step towards the commercially available liquid-nitrogen temperature regime. Particularly, the rigorous two-dimensionality of 1-UC FeSe and FeTe$_{0.5}$Se$_{0.5}$ calls for a fundamentally different mechanism of the potential MZM in 1-UC iron chalcogenides from that proposed for bulk Fe(Te,Se), which relies on the TSSs created through the out-of-plane band inversion[15]. In addition, the 1-UC iron chalcogenides adsorbed by interstitial Fe adatoms also offers overwhelming advantages for studying the ZEBS in the following aspects. 1) The MBE technique utilized for FeSe and FeTe$_{0.5}$Se$_{0.5}$ growth guarantees the highly disorder-free sample quality. 2) The comparatively large SC gaps for 1-UC FeSe and FeTe$_{0.5}$Se$_{0.5}$ protect the ZEBS against perturbations. 3) For future applications, the experimental systems described here *integrate* nearly all the desired ingredients for feasibly realizing and manipulating the MZM-like ZEBS: the unnecessity of external magnetic field for inducing the ZEBS, the ultrashort ZEBS decay length and the technically feasible scanning-tunneling-microscope (STM) manipulation of adatoms further push our experimental systems towards applicable quantum-functionality electronics (Supplementary Part IV).

## Methods

Methods, including statements of data availability and any associated accession codes and references, are available.

## Acknowledgements

The authors acknowledge Ji Feng, X. C. Xie and Qiang-Hua Wang for helpful discussions. **Funding:** This work was financially supported by National Natural Science Foundation of China (No.11888101), National Key R&D Program of China (No. 2018YFA0305604 and No. 2017YFA0303302), National Natural Science Foundation of China (No. 11774008), Strategic Priority Research Program of Chinese Academy of Sciences (No. XDB28000000), Beijing Natural Science Foundation (No. Z180010), and U.S. Department of Energy, Basic Energy Sciences (No. DE-FG02-99ER45747).

## Author contributions

J.W. conceived and instructed the research. C.L. and C.C. prepared the samples. C.L., C.C., Ziqiao W. and Y.L. carried out the STS experiments. X.L. performed the first-principles calculations. C.L., C.C. and S.Y. analyzed the data. C.L., Ziqiang W., J.H. and J.W wrote the manuscript with input from all authors.

## Competing interests

The authors declare no competing interests.

## Additional information

Correspondence and requests for materials should be addressed to J.W.

## Methods

*Sample growth and tunneling experiments.*—The experiments were conducted in an ultrahigh-vacuum (~$2\times10^{-10}$ mbar) MBE-STM combined system (Scienta Omicron). The 0.7%-wt Nb-doped SrTiO$_3$(001) was thermally boiled by 90-°C deionized water for 50 min and chemically etched by 12% HCl solution for 45 min[22]. Then the SrTiO$_3$ was loaded into MBE chamber and pretreated by the Se-flux method[22]. The one-unit-cell (1-UC) FeSe (FeTe$_{0.5}$Se$_{0.5}$) film was grown by coevaporating high-purity Fe (99.994%) and Se (99.999%) [Se (99.999%) & Te (99.999%)] from an e-beam evaporator and a Knudsen cell, respectively, with the SrTiO$_3$ held at 400 °C (340 °C), followed by annealing at 450 °C (380 °C) for 3 h. The Fe adatoms were deposited on 1-UC FeSe and FeTe$_{0.5}$Se$_{0.5}$ surfaces in MBE chamber nominally at ~143–155 K. After annealing and depositing procedures, the 1-UC FeSe and FeTe$_{0.5}$Se$_{0.5}$ films were transferred *in situ* to STM chamber for topography and spectroscopy imaging. A polycrystalline PtIr tip was used throughout the experiments. The topographic images were obtained in a constant-current mode. The scanning tunneling spectra (d$I$/d$V$ vs. $V$) (STS) were acquired by using the standard lock-in technique with a bias modulation typically of 1 mV at 1.7759 kHz. Radio-frequency-noise filters were used to enhance the signal-to-noise ratio. Both the topography and spectroscopy were measured with a bias voltage applied to the tip at 4.2 K unless specified. All the topographic images were post-processed by the SPIP software.

*Theoretical calculations for Fe adatom/1-UC FeSe.*—First-principles calculations were carried out using the Vienna Ab initio simulation package (VASP)[55] with Perdew-Burke-Ernzerhof functional[56]. The LDA+$U$ method[57] with effective $U$ = 0.4 eV on Fe atom was adopted to properly represent the local electronic correlations. The plane-wave cutoff energy was fixed to 400 eV and the crystal structures were relaxed until the atomic forces were smaller than 0.01 eV/Å. A $k$-point mesh of 9 ×9 ×1 centered at Γ point was used in the calculation of single-layer FeSe unit cell. A chequerboard antiferromagnetic spin configuration was adopted[58]. To simulate the Fe adatom, a supercell of 5 ×5 ×1 was constructed. SOC was included for all calculations.

## Data availability

The data that support the findings of this study are available from the corresponding authors upon reasonable request.

Supplementary Information for

# Zero-energy bound states in the high-temperature superconductors at the two-dimensional limit

Chaofei Liu[1†], Cheng Chen[1†], Xiaoqiang Liu[1], Ziqiao Wang[1], Yi Liu[1], Shusen Ye[1], Ziqiang Wang[2], Jiangping Hu[3,4,5,6] & Jian Wang[1,5,6,7]★

[1]International Center for Quantum Materials, School of Physics, Peking University, Beijing 100871, China
[2]Department of Physics, Boston College, Chestnut Hill, MA 02467, USA
[3]Beijing National Laboratory for Condensed Matter Physics & Institute of Physics, Chinese Academy of Sciences, Beijing 100190, China
[4]Kavli Institute of Theoretical Sciences, University of Chinese Academy of Sciences, Beijing 100190, China
[5]Collaborative Innovation Center of Quantum Matter, Beijing 100871, China
[6]CAS Center for Excellence in Topological Quantum Computation, University of Chinese Academy of Sciences, Beijing 100190, China
[7]Beijing Academy of Quantum Information Sciences, West Bld. #3, No. 10 Xibeiwang East Rd., Haidian District, Beijing 100193, China

[†]These authors contributed equally to this work.
★E-mail: jianwangphysics@pku.edu.cn.

**Supplementary Text**

**I. Adsorbate–substrate-interaction–modulated zero-energy bound state**

In our experiments, the Fe adatoms are weakly adsorbed on one-unit-cell (1-UC) FeSe film at the energetically favored hollow sites of surface Se lattice (Fig. 2A and inset of Fig. S1A). Typically, the Fe atoms were deposited with dilute coverage of ~0.002–0.003 monolayer (ML) (e.g., Fig. S1A) (1 ML is defined as the coverage at which Fe adatoms occupy all the hollow sites of surface Se lattice). In statistics, the spins of Fe adatoms are randomly oriented, which locally modulate the magnetic exchange coupling with underlying 1-UC FeSe substrate in different degrees. Thus, the adsorbate–substrate-interaction–dependent Fe-adatom height accordingly appears in a statistical distribution (Fig. S1C). Meanwhile, the energy of the Fe-adatom–induced bound state is similarly distributed (Fig. S1B), yielding a Gaussian profile relatively aligned with adatom-height histogram (Fig. S1, B vs. C). Consequently, the adsorbate–substrate-coupling–tuned bound-state energy for Fe adatom on 1-UC FeSe film is established in statistics. Appropriate exchange interaction between Fe adsorbate and FeSe substrate will pin the bound state at zero energy, which corresponds to a statistical Fe-adatom height of ~62 pm in experiment (Fig. S1C). Particularly, the yielded adatom height for the zero-energy bound state (ZEBS) is relatively large in the Fe-adatom-height histogram (Fig. S1C), suggesting comparably weak adsorbate-substrate interaction. Together with the absence of magnetic-moment extremum at the experimental Fe-adatom height of ~62 pm (Fig. S1D), the adsorbate-substrate coupling for the ZEBS emergence in 1-UC FeSe is evidently below the strong-interaction (unitary) limit.

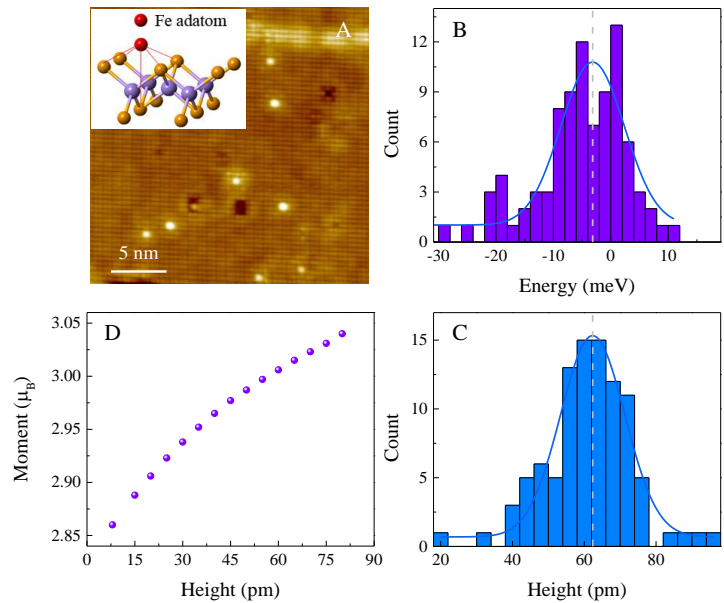

Fig. S1. Statistically correlated bound-state energy and Fe-adatom height for 1-UC FeSe. (A) Topographic image of Fe adatoms on 1-UC FeSe film (25×25 nm$^2$; set point: $V$ = 0.2 V, $I$ = 500 pA). Inset: crystal structure of 1-UC FeSe with a Fe adatom centered above the hollow site of four adjacent Se atoms. (B,C) Bound-state-energy and height statistics for all measured Fe adatoms, both yielding Gaussian distributions (blue curves) with relatively aligned Gaussian-peak positions (dashed lines). The bound state counts with the most intense d$I$/d$V$ in an individual spectrum. (D) Calculated magnetic moment of Fe adatoms on 1-UC FeSe at different heights relative to the upper Se layer (see Methods). $\mu_B$: Bohr magneton.

## II. Additional data of the spatially evolving ZEBS

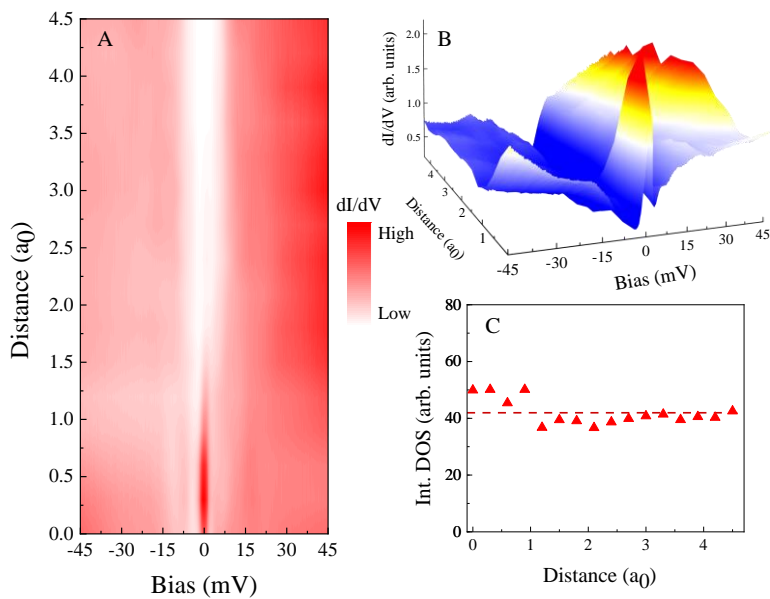

Fig. S2. More analyses of the spatially evolving ZEBS in Fig. 2C. (A,B) 2D and 3D color plots of the spatially resolved tunneling spectra in Fig. 2C. (C) Integrated density of states (DOS) (solid symbols) from −30 to 30 mV for the line spectra in Fig. 2C. The dashed line is the averaged integrated DOS as guide to the eye. The distance in (A)–(C) is defined relative to the Fe-adatom center. (A,B) Set point: $V$ = 0.04 V, $I$ = 2500 pA; modulation: $V_{mod}$ = 1 mV.

## III. Fe-adatom–reconstructed electronic structures

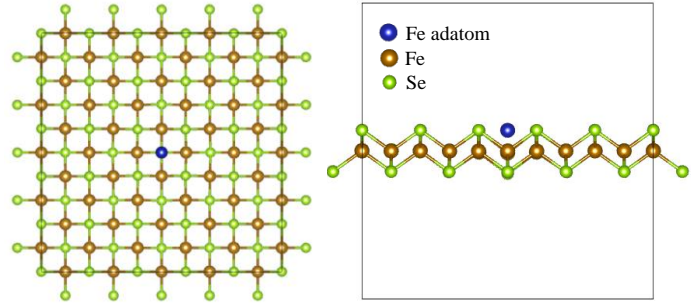

Fig. S3. 5 × 5 × 1 supercell of 1-UC FeSe with Fe adatom (blue) in the center.

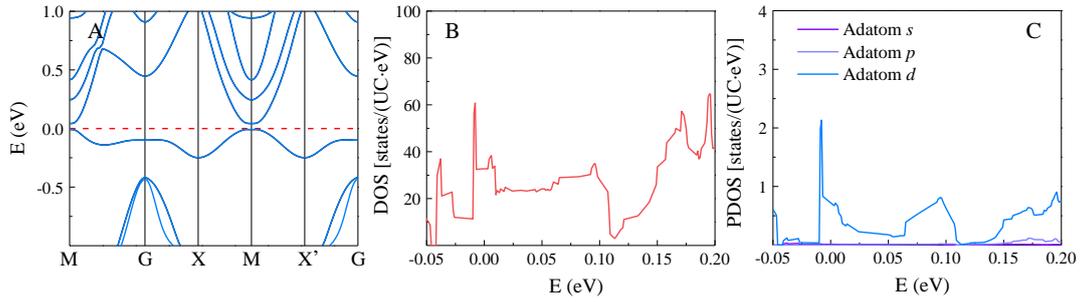

Fig. S4. Electronic structures of 1-UC FeSe reconstructed by Fe adatom. (A) Band structures of free-standing 1-UC FeSe with SOC. The SrTiO$_3$ substrate only introduces electron doping in FeSe which leads to electron pocket[1]. (B) DOS for 5 × 5 × 1 1-UC FeSe supercell with Fe adatom. (C) DOS projected to Fe adatom. For calculation details, see Methods.

**IV. Prospects for the local ZEBS detection in connate high-temperature superconductors**

Compared with the recently prevailing systems hosting the ZEBSs, the 1-UC iron chalcogenides/SrTiO$_3$ *integrate* the desired ingredients for industrially realizing and manipulating ZEBS, which are only partially possessed by individual configurations in previous studies (Table S1; Fig. S5).

(1) *Being connate*. Artificially fabricated Rashba semiconducting nanowires (NWs)[2,3], spin-textured Fe atomic chains[4] and topological-insulator ultrathin films[5] in proximity to Bardeen-Cooper-Schrieffer (BCS) superconductors (Table S1) for pursuing the Majorana zero modes (MZMs) often inevitably introduce the interface unambiguity. The extrinsic complexity of configuration interface will hinder a convincing identification of the ZEBS origin. The connate 1-UC FeSe and FeTe$_{0.5}$Se$_{0.5}$ on SrTiO$_3$ adsorbed by Fe adatoms naturally remove the challenging interface issues.

(2) *High $T_c$ and $T_{ZBCP}$* [$T_c$: critical temperature of the pristine superconducting (SC) components in MZM configurations; $T_{ZBCP}$: existing temperature of experimentally detected zero-bias conductance peak (ZBCP)]. In previous literatures, experiments engineered the ZEBSs mainly based on proposed configurations in proximity to BCS superconductors[2-5] and on bulk iron chalcogenide, Fe(Te,Se)[6,7]. Although bulk Fe(Te,Se) is a nominal high-temperature superconductor, its $T_c$ is generally below 15 K[6]. Therefore, in all previous MZM systems, the generic $T_c$ involved is limited to ~1–10(15) K (Table S1). Further "poisoned" by thermally excited quasiparticles, the MZM-like ZEBS mostly survives at $T_{ZBCP}$ ~0.05–1(4) K (Table S1). As the truly high-$T_c$ (typically 40–65 K[8,9]) superconductors, 1-UC FeSe and FeTe$_{0.5}$Se$_{0.5}$ dramatically push $T_{ZBCP}$ up to 10–13 K and 18–20 K, respectively.

(3) *Unnecessity of external magnetic field*. In SC-proximity–coupled semiconducting NWs and $p_x+ip_y$-wave heterostructures [or bulk Fe(Te,Se)], magnetic field $B$ is required (Table S1) to trigger topological phase transition[10] and to generate magnetic vortices to bound the ZEBSs[11], respectively. The Fe adatoms on 1-UC FeSe and FeTe$_{0.5}$Se$_{0.5}$ films intrinsically induce the ZEBSs under zero external $B$. The unnecessity of magnetic field for inducing the ZEBS in 1-UC iron chalcogenides addresses promising potentials in electronics applications.

Table S1. Summary of the ZEBS-related information in literatures [†for ZBCP-detecting technique; QL: quintuple layer; +: requiring magetic field *B*].

| System | $T_c/T_{ZBCP}$ (K) | $B$ (T) | $\xi$ | Local/Integrated† | Connate/Artificial |
|---|---|---|---|---|---|
| Au/InSb-NW/NbTiN[2] | 12/0.05–0.3 | ≥0.07 | — | Integrated | Artificial |
| Fe chain/Pb(110)[4] | 7/1.4 | 0 | ~1 nm | Local | Artificial |
| IFI/Fe$_{1+x}$(Te,Se)[6] | 12–14/1.5–15 | 0 | 3.5 Å | Local | Connate |
| Point-contacted Cd$_3$As$_2$[12] | ≤7.1/0.28–2.8 | 0 | ~μm | Local | Artificial |
| Magnetic vortex/5-QL Bi$_2$Te$_3$/NbSe$_2$[5] | 7.2/0.03 | + | <10 nm | Local | Artificial |
| Cr-Au/InSb-NW/Al[3] | 1.3/0.02–0.6 | >0.65 | — | Integrated | Artificial |
| Magnetic vortex/FeTe$_{0.55}$Se$_{0.45}$[7] | 14.5/0.55–4.2 | + | 6 nm | Local | Connate |
| Fe adatom/1-UC FeSe | 65/4–13 | 0 | 3.4 Å | Local | Connate |
| Fe adatom/1-UC FeTe$_{0.5}$Se$_{0.5}$ | 62/4–20 | 0 | 6.4 Å | Local | Connate |

(4) *Short decay length*. The ZEBSs in Fe-atomic-chain ends and vortex centers spatially decay by several nanometers[4,5,7] (Table S1). Exceptionally, the decay length of the ZEBSs induced by Fe adatoms on both 1-UC FeSe and FeTe$_{0.5}$Se$_{0.5}$ films is only sub-nanometer (3.4 and 6.4 Å), which is also reported on interstitial iron impurity (IFI) in Fe$_{1+x}$(Te,Se) bulk material[6]. Having an advantage over the IFI/Fe$_{1+x}$(Te,Se) case, the superconductivity in 1-UC FeSe and FeTe$_{0.5}$Se$_{0.5}$ films is regained with a fully gapped and especially

low-bias-DOS–depleted nature instantly away from the Fe adatoms (Fig. 2, C and D; Fig. 4, C and D). The ultralocalized ZEBS and completely recovered SC state within short range in 1-UC iron chalcogenides guarantee the decoupling condition of densely deposited Fe adatoms, making the ZEBS-based high-density information processing (if any) expectable.

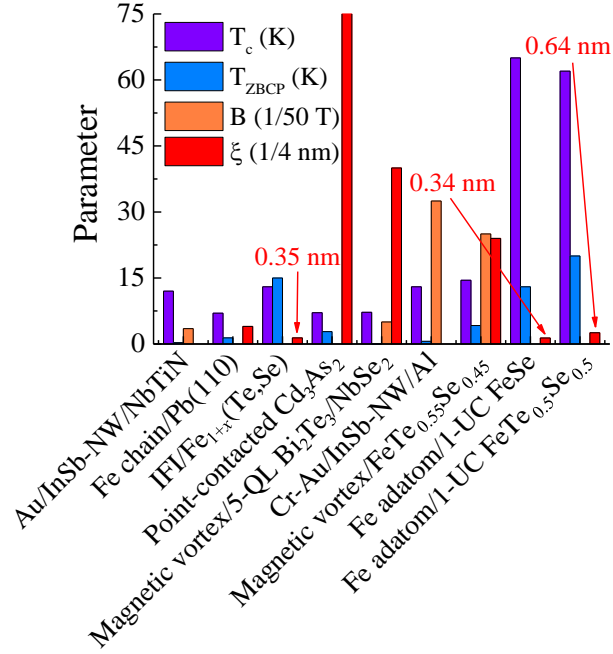

Fig. S5. Histogram of the ZEBS-related parameters in different configurations plotted for direct comparison. Magnetic field $B$ in magnetic vortex/5-QL $Bi_2Te_3$/$NbSe_2$ and magnetic vortex/$FeTe_{0.55}Se_{0.45}$ are drawn by adapting the typical values in original references[5,7]. For concrete parameter values, see Table S1.

(5) *Being locally detected and feasibly manipulated*. The SC-proximity–coupled semiconducting NWs as the MZM platforms are measured by the electrical transport, yielding integrated signals[2,3] (Table S1). Being capable of atomically resolved spectroscopy imaging, the STS utilized in the present experiments highlights the advantage of local ZEBS detection. For state manipulation, the intimate entanglement of the MZM-like ZEBSs at two ends of the vortex line in interfacial $p_x+ip_y$-wave[5] and bulk Fe(Te,Se)[7] systems appears superfluous. In contrast, the paired-ZEBS entangling is nonexistent in the Fe-adatom situation due to the ideal two-dimensionality of 1-UC iron chalcogenides. The technically feasible atomic-manipulation technique by using STM tip further drives the systems of Fe-adatom/1-UC iron chalcogenides towards applicable quantum-functionality electronics.